\documentclass[pra,superscriptaddress,showpacs,amsfonts,amsmath,floatfix,twocolumn]{revtex4} %% two column, final layout

\usepackage{graphicx}
\usepackage{epstopdf}
\usepackage{amssymb}
\usepackage{amsmath}
\usepackage{xspace}
\usepackage{color}

\newcommand{\bra}[1]{\left\langle #1 \right|}
\newcommand{\ket}[1]{\left| #1 \right\rangle}
\newcommand{\aket}[1]{\left| \phi_{#1} \right\rangle}
\newcommand{\abra}[1]{\left\langle \phi_{#1} \right|}

\begin{document}

\title{Thermally generated long-lived quantum correlations for two atoms trapped in fiber-coupled cavities}

\author{Vitalie Eremeev}
\affiliation{Facultad de F\'{i}sica, Pontificia Universidad Cat\'{o}lica de Chile, Casilla 306, Santiago, Chile}

\author{V\'ictor Montenegro}
\affiliation{Facultad de F\'{i}sica, Pontificia Universidad Cat\'{o}lica de Chile, Casilla 306, Santiago, Chile}
\affiliation{Department of Physics and Astronomy, University College London, Gower Street, London WC1E 6BT, United Kingdom}

\author{Miguel Orszag}
\affiliation{Facultad de F\'{i}sica, Pontificia Universidad Cat\'{o}lica de Chile, Casilla 306, Santiago, Chile}

\begin{abstract}
A theoretical model for driving a two qubit system to a stable long-lived entanglement is discussed.
 The entire system is represented by two atoms, initially in ground states and disentangled, each one
 coupled to a separate cavity with the cavities connected by a fiber. The cavities and fiber exchange
  energy with their individual thermal environments. Under these conditions, we apply the theory of
 microscopic master equation developed for the dynamics of the open quantum system. Deriving the density
 operator of the two-qubit system we found that stable long-lived quantum correlations are generated in the presence of thermal
 excitation of the environments. To the best of our knowledge, there is no a similar
 effect observed in a quantum open system described by a generalized microscopic master equation in the
approximation of the cavity quantum electrodynamics.
\end{abstract}


 \maketitle

\section{Introduction}
Entanglement (verschr\"ankung) introduced in physics originally by Schr\"odinger
\cite{Schrodinger} and considered a native feature of the quantum world, is the most outstanding and studied phenomenon to test
 the fundamentals of quantum mechanics, as well as an essential engineering tool for the quantum communications. However,
 entanglement is a  property that is hard to reach technologically and even when achieved, it is a very unstable quantum state,  vulnerable
 under the effects of decoherence, any dissipative process as a result of the coupling to environment. Conventionally these effects
 are considered mainly destructive for entanglement, nevertheless some recent studies of this subject attest results different from the
 common conviction, even appearing as counterintuitive at first glance \cite{{Cirac11}, {Sorensen}, {Memarzadeh}}.

An alternative approach to measure the entire correlations in a quantum system was suggested in Refs. \cite{Vedral, Zurek}. For example,
by using the concepts of mutual information and quantum discord (QD) the quantum correlations may be distinguished from the classical ones.
Further the QD could be compared to the entanglement of formation (E) \cite{Wootters} in order to find if the system is in a quantum inseparable
state (entangled), or in a separable state with quantum correlations, sucha as QD \cite{Luo, Alber, Lu, Fanchini}. Such an analysis is considered in this paper.

The inclusion of the interaction of the system with the environment plays an important role in physics, implying a more realistic picture because
the dissipation is always present in the real devices. In the proposed study we deal with atoms, cavities and a fiber in the framework of the
physical model suggested in Ref. \cite{Cirac97}, which attracted a high interest for quantum information applications and subsequently
discussed details from different aspects \cite{Pellizzari, Mancini, Serafini, Zheng}. As a basic model, we consider the one recently analyzed in Ref.
\cite{Mont} and extend the calculations for a very special case, i.e. when the atoms are initially disentangled and in the ground states while the
fields are in vacuum states and coupled to the reservoirs at finite temperatures. The entire system is considered open because of the leakage of
the electromagnetic field from the cavities and fiber into their own thermal baths. Therefore, we ask ourselves the following question: Is
it possible to generate atomic quantum correlations by the processes of absorption and
 exchanging excitations with the thermal reservoirs? In the following we present the model and detailed analysis in search of an answer.

%\section{Model}
\section{Model}
\begin{figure}[t]
\includegraphics[width=0.95 \columnwidth]{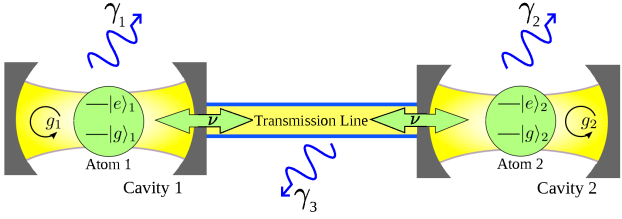}
\caption{Two atoms trapped in distant coupled cavities. The cavities and transmission line exchange the energy at the rates
  $\gamma_1$, $\gamma_2$ and $\gamma_3$ with their baths having the temperatures $T_1$, $T_2$ and $T_3$, respectively.}
  %%$g_i$ and $\nu$ are the atom-cavity and cavity-fiber coupling constants.
\label{fig1}
\end{figure}

We present here the model schematically shown in Fig.\ref{fig1} and recall the basic equations that lead us
to the effect we are looking for. Hence, one considers two qubits (two-level atoms) interacting with two different and distant cavities, coupled
by a transmission line (e.g., fiber, waveguide). For simplicity we consider the short fiber limit: only one mode of the fiber
 interacts with the cavity modes \cite{Serafini}.

Now, let us define a given state of the whole system by using the notation:
$\ket{i}=\ket{A_1}\otimes\ket{A_2}\otimes\ket{C_1}\otimes\ket{C_2}\otimes\ket{F} \equiv \ket{A_1A_2C_1C_2F}$, where
${A}_{j=1,2}$ correspond to the atomic states, that can be $e(g)$ for excited(ground) state, while ${C}_{j=1,2}$ are the
 cavity states, and ${F}$ corresponds to the state of the fiber. Both ${C}_{j=1,2}$ and $F$ describe a $0$ or $1$
 photon state. The Hamiltonian of the composite system under the rotating-wave approximation (RWA) reads (with $\hbar = 1$)
\begin{eqnarray}
H_s &=& \omega_a a_3^{\dag} a_3+\sum_{j=1}^2  \left( \omega_a S_{j,z}+\omega_0 a^{\dag}_j a_j \right) \nonumber \\
&+& \sum_{j=1}^2 \left( g_j S^+_j a_j + \nu a_3 a^{\dag}_j + H.c.\right),
\label{Ham}
\end{eqnarray}
where $a_3$ is the boson operator defining the fiber mode, $a_1(a_2)$ is the boson operator for the cavity 1(2);
 $\omega_0$ and $\omega_a$ are the cavity and the atomic (fiber as well) frequencies, respectively;  $g_j$ ($\nu$) the atom-cavity
 (fiber-cavity) coupling constants; and $S_{z}$, $S^{\pm}$ are the usual atomic inversion and ladder operators, respectively.
 %Fig1
 The model is studied under the assumption of a single excitation in the system of atoms and fields, and using the above mentioned
notation, the state-basis of the system becomes: $\ket{1} = \ket{eg000},  \ket{2} = \ket{gg100}, \ket{3} = \ket{gg001},
 \ket{4} =\ket{gg010},  \ket{5} = \ket{ge000},  \ket{6} = \ket{gg000}$, where the last vector is required by the existence
 of the excitation's leakage to the reservoirs. Hence, it is straightforward to bring the Hamiltonian $H_s$ in Eq. (\ref{Ham})
 to a matrix representation in the state-basis \cite{Mont}. %(for more details see Eq. (4))

To simulate the dynamics of the given system, one considers the approach of the microscopic master equation (MME), developed in Refs. \cite{Scala,
Breuer} in order to describe the system-reservoir interactions by a Markovian master equation. This description considers jumps between
eigenstates of the system Hamiltonian rather than the eigenstates of the field-free subsystems, which is the case in many approaches employed in
quantum optics. Therefore, we assume that the system of interest, i.e. the atoms, cavities and fiber are parts of a larger system, composed by a
collection of quantum harmonic oscillators in thermal equilibrium.
 The external environment represents the part of the entire closed system other than the system of interest. Between each element
 of the system and its own bath one may identify different kind of dissipation channels. In cavity quantum electrodynamics (CQED) the main source of
  dissipation originates from the leakage of the cavity photons due to the imperfect reflectivity of the cavity mirrors. A second
  source of dissipation corresponds to the spontaneous emission of photons by the atom, however this kind of loss we consider small and neglect
  in the model. Following the common procedures \cite{Scala, Breuer}, one obtains the MME for the system's reduced density operator $\rho(t)$
\begin{equation}
\frac {\partial \rho}{\partial t}=-i\left[ H_s,\rho \right]+\mathcal{L}(\bar{\omega}) \rho+\mathcal{L}(-\bar{\omega}) \rho,
\label{MME}
\end{equation}
where the dissipation terms are defined as follows (with $\bar{\omega}>0$)
\begin{equation}
\mathcal{L}(\bar{\omega}) \rho = \sum_{j=1}^3 \gamma_j(\bar{\omega})\bigg(A_j(\bar{\omega})\rho A_j^ \dag (\bar{\omega})- \frac{1}{2} \left[ A_j^
\dag (\bar{\omega}) A_j(\bar{\omega}),\rho \right] _{+} \bigg). \nonumber
\label{Lind}
\end{equation}
In the above equations the following definitions are considered: $A_j(\bar{\omega}) = \sum_{\bar{\omega}_{\alpha, \beta}} \aket{\alpha}
\abra{\alpha} (a_j + a_j^{\dag}) \aket{\beta}\abra{\beta}$ fulfilling the properties $A_j(-\bar{\omega})= A_j^{\dag}(\bar{\omega})$,
where $\bar{\omega}_{\alpha, \beta} = \Omega_{\beta} - \Omega_{\alpha}$ with $\Omega_k$ as an eigenvalue of Hamiltonian $H_s$ and its
 corresponding eigenvector $\aket{k}$, denoting the \textit{k}-th dressed-state. We should point out that the eigenfrequencies of
 Hamiltonian $H_s$ are chosen in order to satisfy the following inequality $\Omega_6< \Omega_5< \Omega_4< \Omega_3< \Omega_2< \Omega_1$.
  Further in Eq. (\ref{MME}) one may use the so-called Kubo-Martin-Schwinger (KMS) condition \cite{Breuer}, which gives a relation for
  the damping constants $ \gamma_j(-\bar{\omega}) = \mathrm{exp}\left(-\bar{\omega}/ T_j\right)\gamma_j(\bar{\omega})$, where $T_j$ are the
  reservoir temperatures in the corresponding unit. The KMS condition ensures that the system tends to a thermal equilibrium for $t \to \infty$.

In order to solve Eq. (\ref{MME}) one may use a kind of formal solution, because in the most general case there is no an analytic solution for the
eigenvalue equation based on Hamiltonian $H_s$. Once having the operators $A_j(\bar{\omega}_{\alpha \beta})$, it is easy to write Eq.
(\ref{MME}) for the density operator $\rho(t)$ decomposed in the eigenstates basis, $\bra{\phi_m}\rho(t)\ket{\phi_n} = \rho_{mn}$, and  we get
\begin{eqnarray}
\dot{\rho}_{mn} = - i \bar{\omega}_{n,m} \rho_{mn} + \sum_{k=1}^5 \frac{\gamma_{k \to6}}{2} \big( 2\delta_{m6}\delta_{6n}
\rho_{kk} - \delta_{mk}\rho_{kn}
\nonumber \\
- \delta_{kn} \rho_{mk} \big)+ \sum_{k=1}^5 \frac{\gamma_{6\to k}}{2} \big( 2\delta_{mk}\delta_{kn}  \rho_{66} - \delta_{m6}\rho_{6n} -
\delta_{6n}\rho_{m6}  \big)
 \label{rhosys}
\end{eqnarray}
Here $\delta_{mn}$ is the Kronecker $\delta$; the physical meaning of the damping coefficients $\gamma_{k \to 6}$ and $\gamma_{6 \to k}$ refers to
the rates of the transitions between the eigenfrequencies $\Omega_k$ downward and upward, respectively, defined as follows $\gamma_{k \to
6}=\sum_{j=\{1,2,3\}} c_i^2\gamma_j \left[\langle n(\bar{\omega}_{6,k})\rangle_{T_j} + 1\right] $ (similar to Eq.(13) in Ref. \cite{Mont} redefined for finite temperature) and $\gamma_{6 \to k}$ results from the KMS
condition, where $c_i$ (with the integer $i$ varying from 1 to 25) are the elements of the matrix for the transformation from the states $\{\ket{1}, ... , \ket{6}\}$ to the states $\{
\aket{1}, ... , \aket{6} \}$ (see Eq. (14) and Appendix A in \cite{Mont}). Here $\langle n(\bar{\omega}_{\alpha, \beta})\rangle_{T_j} = \left
( \mathrm{e}^{(\Omega_\beta- \Omega_\alpha) / T_j} - 1\right )^{-1}$ corresponds to the average number of the thermal photons. The damping
coefficients play the central role in our model because their dependence on the temperature of the reservoirs implies a complex exchange mechanism
between the elements of the system and the baths. Therefore, in the presence of the temperature we solve numerically the coupled system of the
first-order differential equations (\ref{rhosys}) and compute the evolution of entanglement considering the atom-field system in the initial unexcited state
$\ket{gg000}$.
 %Fig2
\begin{figure*} [t]
\includegraphics[width=0.95 \textwidth]{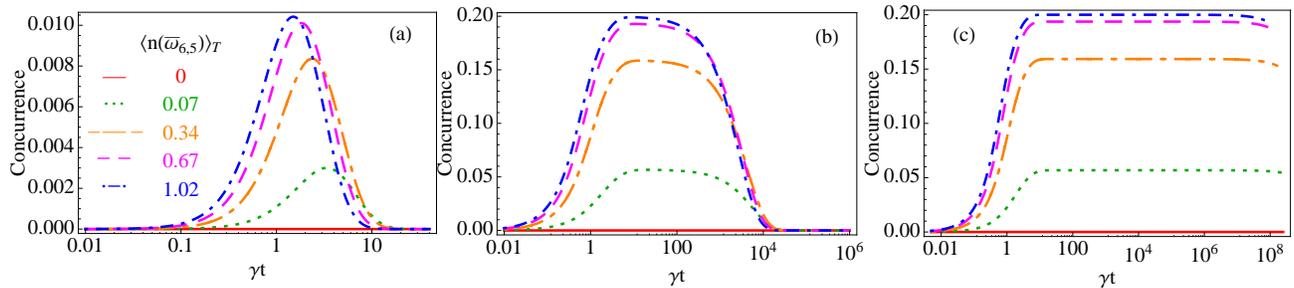}
\caption{Evolution of the concurrence for $g=\nu=5\gamma$ and different atom-cavity
detunings: (a) $\Delta =0$ , (b) $\Delta=10^{-4}\omega_a $ and (c) $\Delta=0.1 \omega_a $. The baths have the same temperature with the average
number of thermal photons given by $\langle n(\bar{\omega}_{6, 5})\rangle_{T}$. The axis of the dimensionless time, $\gamma t$, is in a
logarithmic scale.}
\label{fig2}
\end{figure*}

In the next section we present the calculations of the quantum correlations depending on the system characteristics, such as atom-cavity detuning, coupling constants and thermal reservoirs.
In order to compute the atomic entanglement, we need to perform a measurement of the cavities-fiber field with a state $\ket{000} =
\ket{0}_{C1}\otimes \ket{0}_{C2} \otimes \ket{0}_{F}$. The feasibility of such a measurement is also discussed.

% \section{Measuring the quantum correlations}
 \section{Measuring the quantum correlations}
 \subsection{Entanglement}
 Once projected on the state $\ket{000}$ of the field subspace we find that the reduced atomic density matrix in the two-qubit basis $\{ \ket{gg}, \ket{ge}, \ket{eg}, \ket{ee} \}$ preserves during the time evolution a X-form structure
 \begin{equation}
\tilde {\rho}(t)=\begin{pmatrix}  \tilde {\rho}_{11} &0&0&0\\ 0& \tilde {\rho}_{22}& \tilde {\rho}_{23}&0\\ 0& \tilde {\rho}_{32}& \tilde {\rho}_{33}&0\\ 0&0&0&0 \end{pmatrix},
\label{xform}
\end{equation}
with the atoms initially in ground state [i.e. $\tilde {\rho}_{11} (0)=1$]. The entanglement measured by the concurrence \cite{Wootters} could be easily computed \cite{Mont} and gives $C(t)=2\mid \tilde {\rho}_{23}/(\tilde {\rho}_{11}+\tilde {\rho}_{22}+\tilde {\rho}_{33}) \mid$. A more particular form of the density matrix (4) may result in the case of interchanging the undistinguished qubits in equivalent cavities (i.e. for $g_1=g_2$, $\gamma_1=\gamma_2$ and $T_1=T_2$).  

In the following, we are mainly interested in studying the evolution of atomic entanglement, the concurrence ($C$), as a function of
 the temperatures of the thermal baths. The system under consideration refers to the atoms with long radiative lifetimes, each coupled to its own
cavity. These two cavities are connected by a fiber with the damping rates $\gamma_1=\gamma_2=\gamma_3 \equiv \gamma=2 \pi$ MHz,
respectively, which are within the current technology \cite{Serafini}. The transition frequency of the atom is chosen to be mid-infrared (MIR), i.e.
$\omega_a/2 \pi=4$THz and hence, for experimental purposes the coupling between the distant cavities can be realized by using the modern resources
of IR fiber optics, e. g. hollow glass waveguides \cite{Harrington}, plastic fibers \cite{Chen}, etc.
We choose the range of MIR frequencies in order to limit the thermal reservoir only up to room temperature (300K), which corresponds to a thermal
photon. The values of the coupling constants and the atom-cavity  detuning will be varied in order to search the optimal result. We must mention
here that to satisfy the RWA we should have
 $2g\gg \gamma_{max}(\bar{\omega})$ \cite{Scala}. Satisfying this condition we start with the case  $g_1=g_2\equiv g=\nu=5\gamma$,
 considering all the reservoirs at the same temperature, $T$, and study how the atomic entanglement evolves as a function of the atom-cavity
 detuning, $\Delta$.
The result is shown in Fig. \ref{fig2} from which we conclude that the atom-cavity detuning facilitate in this case the
 generation of a quasi-stationary atomic entanglement and for $\Delta=0.1\omega_a$ the system reaches a long-lived entanglement state. Of course,
 in the asymptotic limit the concurrence will vanish and the atoms eventually disentangle themselves due to the damping action of the reservoirs.
 The maximal value of the concurrence of $\sim$0.2 corresponds to  the bath's temperature about 300 K, that is about one thermal excitation (consistent with the single-excitation approximation in the model) for
 the given frequency  $\omega_a$, so that $k_B T/\hbar \omega_a \simeq 1.5$.
 %Fig3
\begin{figure} [b]
\includegraphics[width=0.9 \columnwidth]{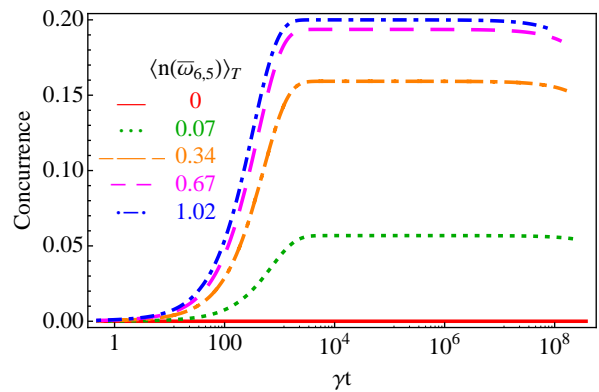}
\caption{Concurrence for $\Delta=0.1 \omega_a $, $g=5\gamma$ and $ \nu=100\gamma.$}
\label{fig3}
\end{figure}
%Fig4
 \begin{figure*} [t]
 \includegraphics[width=0.95 \textwidth]{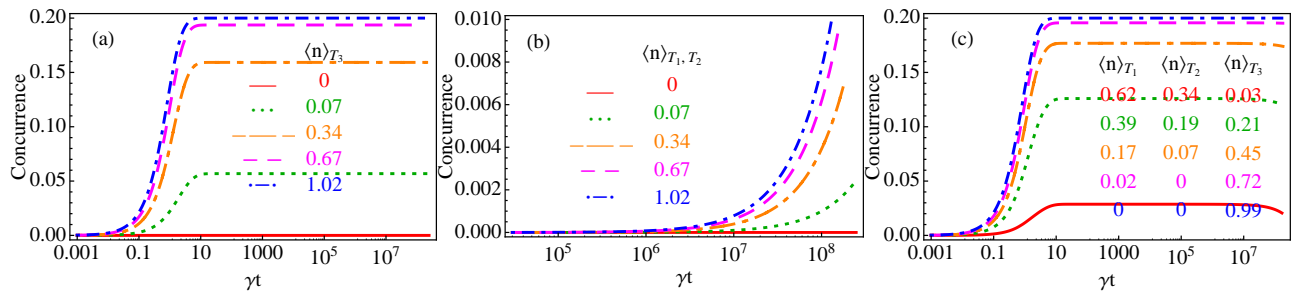}
 \caption{Evolution of the concurrence for arbitrary baths temperatures, (a) $T_1=T_2=0$ and varying the fiber's bath temperature,
 (b) $T_3=0$ and varying equally the cavities' bath temperatures, and (c) varying differently all the temperatures.
  The rest of the  parameters are the same as in Fig. \ref{fig2}(c).}
 \label{figconctemp}
 \end{figure*}
In order to find the optimal relation between the coupling constants and damping rate we did the calculations for different situations
as follows: (i) $ g=\nu=100\gamma$, (ii) $ g=5\gamma$ and $ \nu=100\gamma$, (iii) $ g=10\gamma$ and $ \nu=\gamma$.
For example, we present the case (ii) in Fig. \ref{fig3}, from which we see that the concurrence gets the same maximal value as in
the previous case Fig. \ref{fig2}(c), but it takes a longer time for the quasi stationary entanglement to reach its plateau.
The rest of the cases give worse results.

Now, let us analyze a more general situation, when all the independent baths have different temperatures. After performing the computations,
we found an interesting effect that only the thermal bath of the fiber plays an important role in the generation of entanglement in the system,
while the thermal baths of the cavities generate very little entanglement. This situation is represented in Fig. \ref{figconctemp}. Therefore,
 after analyzing all the calculations at the given circumstances, we come to the conclusion that the case represented in Fig. \ref{fig2}(c) corresponds to the optimal
 one for the generation of entanglement.
 %Fig5
\begin{figure} [b]
 \includegraphics[width=0.9 \columnwidth]{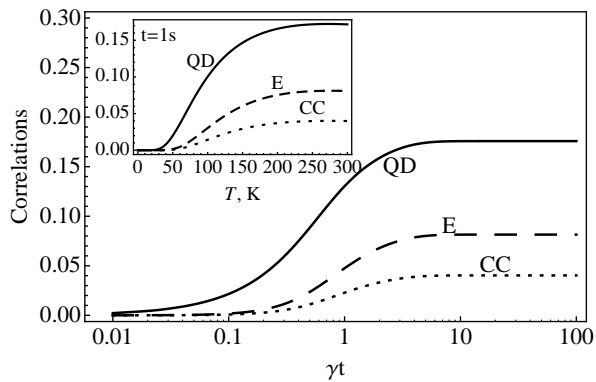}
\caption{Evolution of the quantum discord (QD), entanglement of formation (E) and classical correlations (CC) for one thermal
 excitation and the parameters chosen as in Fig. \ref{fig2}(c). The inset represents the same quantities as a function of the  temperatures
 of the reservoirs  calculated for a late time, $t=1$s.}
\label{figcorrel}
\end{figure}

\subsection{Quantum Discord}
Alternatively to entanglement, the quantum correlations can be also quantified by the quantum discord \cite{Vedral, Zurek, Luo, Alber, Lu,
Fanchini}. Since in our case the two-qubit density matrix has a simplified $X$ form (\ref{xform}), one can easily compute the quantum and classical
correlations in the system by using a particular case for the algorithm discussed in \cite{Alber}. Even if some recent studies, such as \cite{Lu}, found
that the analytic approach of Ref. \cite{Alber} could not be considered as a general one, in our case the computation of QD may follow this procedure
without some divergences of the minimization approach. In the framework of the algorithm and notations used in \cite{Alber}, we have to optimize
QD just by changing the parameters $(k,l)$ in the range $(0,1)$ and found easily the condition of the resultant minimum for $(k, l)$=1/2. We have
also compared the calculations with the approach proposed in \cite{Fanchini}, by using Eq. (6) of
 the later and obtained exactly the same result. Hence, we observe in Fig. \ref{figcorrel}
the time evolution of the QD similar to that of entanglement, but the initial growth is steeper in the discord, which implies
the appearance of the quantum correlations in the system prior to the entanglement \cite{Davidovich}. For a better illustration of the
thermal effect under discussion, in the inset is shown the temperature dependence of the steady values (flat time plateau) of the quantum
and classical correlations.

\subsection{Experimental hint}
In the following, we discuss the tasks important for an experimental realization of the ideas discussed here.
In our opinion, the most difficult is to realize a quantum non-demolition (QND) measurement of the photon
states in the fiber-coupled cavities. However, nowadays there exist technological possibilities to realize experiments on QND photon counting,
 attaining single-quantum resolution, performed with optical or microwave photons \cite{Guerlin} (for an exhaustive review see
Ref.  \cite {Grangier}). In the experiment discussed in Ref. \cite{Guerlin} the cavity mode was coupled to Rydberg atoms or superconducting
  junctions and the QND method is based on the detection of the dispersive phase shift produced by the field on the wave function of non-resonant atoms
  crossing the cavity. This shift can be measured by atomic interferometry, using the Ramsey separated-oscilatory-field method. The advantages
   of QND experiments in radiometry and in particular applied for IR photons are suggested in \cite{Castelleto}.

\begin{figure} [t]
 \includegraphics[width=0.9 \columnwidth]{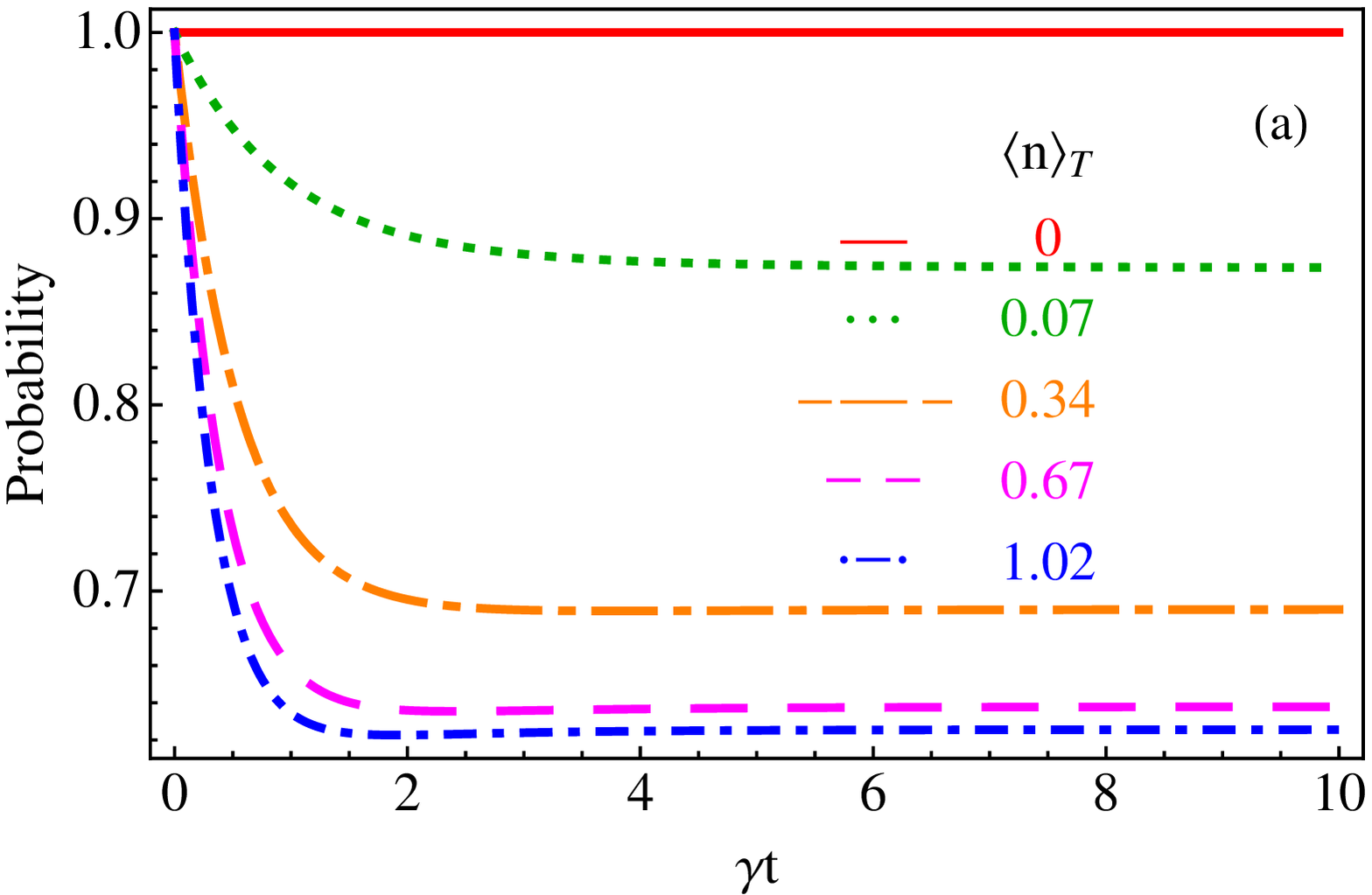}
  \includegraphics[width=0.9 \columnwidth]{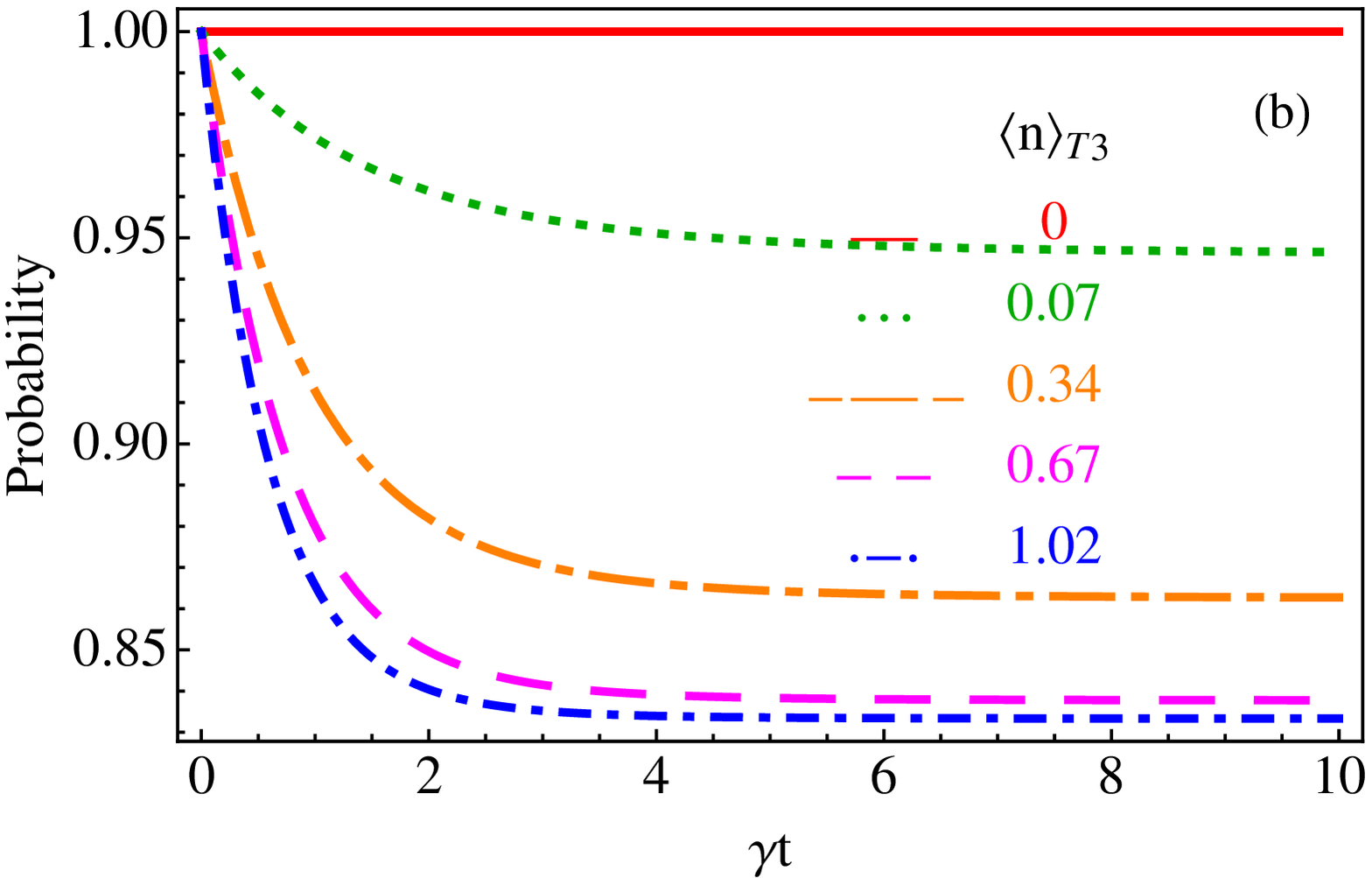}
\caption{Probability of finding simultaneously the fiber and the cavities in a vacuum state by engineering of the thermal baths as follows: (a) all the baths have the same temperature $T$, (b) varying the fiber's bath temperature $T_3$, while $T_1=T_2=0$. The rest of the parameters are the same as in Fig. \ref{fig2}(c).}
\label{figQND}
\end{figure}

In order to simulate a measurement on the fiber-cavity subsystem one may compute the field density operator and therefore monitor the
 probability of the field state. As we are interested in preserving the field in the vacuum state, i.e. $\rho_{fib-cav}(t)=\ket{000}\bra{000}$,
 one tests the probability of this state during the temporal evolution of the system. The dynamics of this probability for different schemes
 of engineering of the thermal reservoirs is shown in Fig. \ref{figQND}. Based on these results we conclude that the success to find the
 fiber-cavity field in a vacuum state after the measurement strongly depends on the managing of the thermal reservoirs. Hence, from this
 point of view, a more efficient variant to drive the qubits to long-lived quantum correlations is to increase the fiber's bath temperature
  while the baths of the cavities are maintained at the lowest possible temperature.
 %Fig6

%\section{Concluding remarks}
\section{Concluding remarks}
In this study we show a very interesting effect that the long-lived quantum correlations between the atoms trapped in separate cavities can be
generated by the dissipative coupling to the thermal baths. This is an example that could give us insight into the effects of the
system-environment exchange versus the quantum correlations. From the analysis of the obtained results, mainly Fig. \ref{figconctemp} and
\ref{figQND}, we conclude that the entanglement can be optimized by engineering the thermal bath of the fiber rather than the baths of each
cavity, hence suggesting that the quasilocal manipulations produce little effect on the generation of entanglement. Furthermore, we found that
our system evidences quantum correlations quantified by QD prior to the appearance of the entanglement (Fig. \ref{figcorrel}). Summarizing, the model discussed here can
be implemented as a QND measurement on the cavity-fiber fields with a high success probability (Fig. \ref{figQND}). This is an example of a system where quantum
correlations are only driven by thermal excitations and can be of interest as an alternative method for protection and generation of quantum
correlations.

\acknowledgments
M.O. acknowledges support from Fondecyt, Grant No. 1100039, and V.E. gratefully acknowledges postdoctoral support from the Physics Faculty, Pontificia Universidad Cat\'{o}lica de Chile.

\end{document}